\documentclass[aps,prl,floatfix,twocolumn]{revtex4-1}

\usepackage{amsfonts,amsmath,amssymb}
\usepackage{textcomp}
\usepackage{bm}
\usepackage{upgreek}
\usepackage[dvips]{graphicx}
\usepackage[latin1]{inputenc}
\usepackage{natbib}

\renewcommand{\vec}[1]{\bm{#1}}
\newcommand{\ee}{\mathrm{e}}
\newcommand{\ii}{\mathrm{i}}

\newcommand{\fracsmall}[2]{\mbox{$\frac{#1}{#2}$}}

\newcommand{\aaa}{\vec{a}}

\newcommand{\kkk}{\vec{k}}

\newcommand{\rrr}{\vec{r}}

\newcommand{\calA}{\mathcal{A}}
\newcommand{\calQ}{\mathcal{Q}}

\newcommand{\calR}{\mathcal{R}}
\newcommand{\calT}{\mathcal{T}}

\newcommand{\beq}[1]{\begin{equation} \eqlab{#1}}
\newcommand{\eeq}{\end{equation}}
\newcommand{\bsub}{\begin{subequations}}
\newcommand{\esub}{\end{subequations}}
\def\bal#1\eal{\begin{align}#1\end{align}}
\def\bsubal#1\esubal{\bsub \begin{align}#1\end{align} \esub}

\newcommand{\eqlab}[1]{\label{eq:#1}}
\renewcommand{\eqref}[1]{Eq.~(\ref{eq:#1})}

\newcommand{\figref}[1]{Fig.~\ref{fig:#1}}

\begin{document}

\title{Theoretical analysis of a dual-probe scanning tunneling microscope setup on graphene}

\author{Mikkel Settnes}\email[]{mikse@nanotech.dtu.dk}
\author{Stephen R. Power}
\author{Dirch H. Petersen}
\author{Antti-Pekka Jauho}
\affiliation{Center for Nanostructured Graphene (CNG), Department of Micro- and Nanotechnology Engineering, Technical University of Denmark, DK-2800 Kgs. Lyngby, Denmark}

\date{\today}

\begin{abstract}
Experimental advances allow for the inclusion of multiple probes to measure the transport properties of a sample surface. 
We develop a theory of dual-probe scanning tunnelling microscopy using a Green's Function formalism, and apply it to graphene. 
Sampling the local conduction properties at finite length scales yields real space conductance maps which show anisotropy for pristine graphene systems and quantum interference effects in the presence of isolated impurities.
The spectral signatures of the Fourier transform of real space conductance maps include characteristics that can be related to different scattering processes.
We compute the conductance maps of graphene systems with different edge geometries or height fluctuations to determine the effects of non-ideal graphene samples on dual-probe measurements.

\end{abstract}

\maketitle
Local scattering centers such as impurities, defects and substrate inhomogeneities limit the theoretically high mobility of graphene \cite{CastroNeto2009,Chen2008,Hwang2007}. 
Improved sample preparation and specialized substrates have improved the quality of graphene electronics \cite{Dean2010} such that even a single scatterer can influence the whole device and perhaps render it useful for, e.g. sensing applications \cite{Schedin2007,Brar2010}. 
A detailed understanding of the influence of such defects on electronic properties is necessary in order to exploit or avoid their influence \cite{Peres2006,Peres2010}.

Information about single scatterers can be obtained via scanning tunnelling microscopy (STM), yielding direct information about the local density of states (LDOS). Previously the LDOS of graphene has been studied, both experimentally and theoretically, in the presence of defects \cite{Bena2008,Simon2009,Amara2007,Mallet2007,Rutter2007,Peres2009,Cheianov2006,Bacsi2010}, edges \cite{Xue2012,Yang2010,Park2011,Barone2006, Mason2013}, constrictions \cite{Bergvall2013} and charge puddle formation caused by trapped molecules \cite{Deshpande2009,Zhang2009}.

However, in many contexts one is interested in how the local electronic transport properties, and not just the LDOS, vary along the sample. To this aim multi-probe STM has been used to characterize a wide range of systems, including carbon nanotubes \cite{Nakayama2012}, Si-nanowires \cite{Qin2012,Cherepanov2012}, two-dimensional thin films \cite{Bannani2008} and graphene \cite{Cherepanov2012,Ji2012,Sutter2008}. This technique analyzes nanoscale features on surfaces without the need to fabricate invasive contacts into the sample \cite{Sutter2008,Wang2013,Ji2012,Buron2012}. Graphene is especially interesting as it is intrinsically two-dimensional and we thus probe the material properties by measuring the surface. Furthermore graphene has a long inelastic mean free path \cite{Borunda2013,Mayorov2011,Berger2006,Bolotin2008,Rickhaus2013}, enabling the possibility of placing two STM tips within a length scale at which interference effects are not washed out by dephasing \cite{Nakayama2012,Baringhaus2013,Eder2013,Rickhaus2013}.

\begin{figure}[tb]
\centering
\includegraphics[width=0.75\columnwidth]
{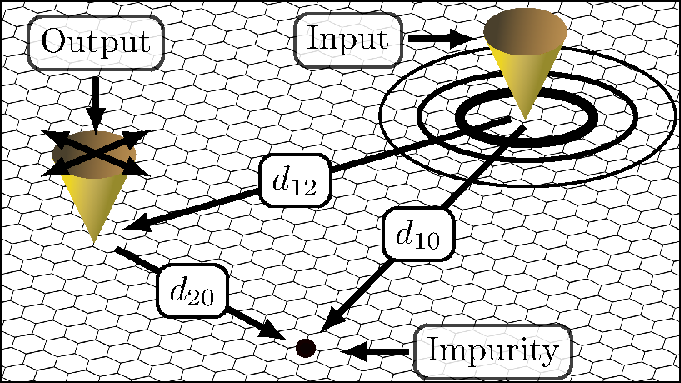}
\caption[]{\small Schematic overview of a dual-probe STM setup. Current input/output probes and an impurity on site $0$ are indicated together with their relative separations.} \label{fig:sketchT0}
\end{figure}

In this Letter we consider such quantum interferences as we present a theoretical analysis of the dual-probe STM setup as sketched in \figref{sketchT0}. 
The methodology and analysis is described for pristine graphene sheets and vacancies, but is completely general and can be easily extended to other systems.
Applications to graphene systems with edges or height fluctuations are presented as examples.

{\it Methods.}
-- In nonequilibrium Green's function formalism (NEGF) semi-infinite leads are coupled to a finite device region \cite{DattaBook,AnttiBook}. 
We instead consider an {\it infinite} two-dimensional device connected to one fixed and one scanning STM probe as in \figref{sketchT0} so that conventional recursive methods are not directly applicable, and an alternative approach must be used. Although we consider graphene in this work, the method is applicable to other surfaces by using the relevant Green's Function (GF) in the following derivations. For pristine graphene in the nearest neighbour tight-binding model, the real-space single-particle equilibrium GF is given by \cite{Power2011} 
\bal
g^0_{ij}(z) = \frac{1}{\Omega_{BZ}}\int\mathrm{d}^2\kkk \frac{\ee^{\ii \kkk\cdot (\rrr_j-\rrr_i)}}{z^2 - t^2|f(\kkk)|^2} \begin{pmatrix}
z & tf(\kkk) \\
t f^*(\kkk) & z
\end{pmatrix} , \eqlab{g0}
\eal
where $z = E +\ii 0^+$ is the energy, $\Omega_{BZ}$ is the area of the first Brillouin zone, 
$\rrr_{i} = m_{i}\aaa_1+n_{i} \aaa_2$ (with $m_i$ and $n_i$ integers) is the position of site $i$, $\aaa_1$ and $\aaa_2$  are the graphene lattice vectors, and $f(\kkk) = 1 + \ee^{\ii \kkk \cdot\aaa_1} + \ee^{\ii \kkk \cdot\aaa_2}$. The carbon-carbon hopping integral is $t\approx -2.7$ eV \cite{Reich2002}. 

The zero-temperature conductance is given by the Landauer formula $\fracsmall{2e^2}{h} \calT_{12}$ \cite{DattaBook}, where the transmission coefficient between the two probes is
\bal
\calT_{12}(E) = \mathrm{Tr}\big[G(E)\Gamma_2(E) G^\dagger(E) \Gamma_1(E)\big], \eqlab{trans}
\eal
$\Gamma_{i}(E)$ ($i=1,2$) is the coupling to the probes and $G/G^\dagger$ is the retarded/advanced Green's function of the sample including probe effects. 

Experimental STM tips have finite radii of curvature, limiting the resolution due to couplings with multiple lattice sites. We employ the Tersoff-Hamann approach \cite{Meunier1998,Fukuda2007,Nakanishi2010} to describe a structureless tip with only the end orbital of a linear atomic chain coupled to the sample. The DOS of the chain is constant in the considered energy range. The coupling between the tip and a nearby lattice site $i$ is angle dependent and decays exponentially with separation \footnote{The coupling is calculated using\cite{Meunier1998,Amara2007}, $t_i = t_0 w_i \ee^{-d_i/\lambda}\cos\big(\theta_i\big)$ where $w_i = \ee^{-ad_i^2}/\sum_{j} \ee^{-ad_j^2}$, $\theta_i$ is the angle between the tip apex and site $i$, $\lambda=0.85\AA$, $a=0.6 \AA^{-2}$. $t_0$ is a scaling factor which we set to $t_0=10t$.}. The results presented below are in broad agreement with test calculations performed for more realistic tips, where a predictable smearing of the shorter range features occurs.

%

\begin{figure}[tb]
\centering
\includegraphics[width=0.90\columnwidth]
{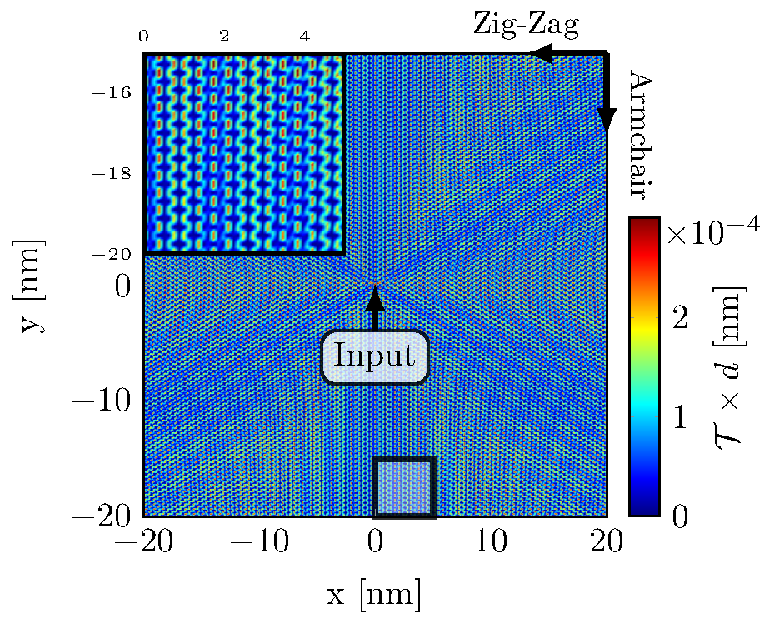}
\caption[]{The conductance map for pristine graphene with $E_F=0.5|t|$. The fixed input probe is at the origin, and the map represents the conductance between the probes as a function of scanning probe position. The conductance has been multiplied by the inter-probe distance $d_{12}$ to compensate for a geometric decay,  see \eqref{SPA}. The inset is a magnification of the boxed area.} \label{fig:T0}
\end{figure}

{\it Pristine Graphene.}
-- The transmission $\calT_{12}$ is obtained from \eqref{trans} using a numerical evaluation of \eqref{g0}. 
The resulting map is shown in \figref{T0} for $E_F = 0.5|t|$. 
Other Fermi energies show similar qualitative behaviour, but lower $E_F$ values require a larger scan area to obtain the same number of oscillation periods.
In armchair directions a constant $\calT_{12}\times d_{12}$ transmission is observed, while oscillations occur for zigzag directions.
The results are not very sensitive to the exact position of the stationary probe, with the exponential coupling generally ensuring that the probe primarily couples to a single site.

To qualitatively understand the different behaviour for the two high symmetry directions, we exploit the fact that \eqref{g0} can be approximated analytically for separations above a few lattice spacings using the stationary phase approximation (SPA)\cite{Power2011}. 
The GF can thus be written for the armchair and zigzag directions, respectively, as 
\bal
g^{0,ac}_{ij} = \frac{\calA(E)\ee^{\ii \calQ(E) d_{ij}}}{\sqrt{d_{ij}}},\quad g^{0,zz}_{ij} = \sum_{\eta=\pm}\frac{\calA^\eta(E)\ee^{\ii \calQ^\eta(E) d_{ij}}}{\sqrt{d_{ij}}},\eqlab{SPA}
\eal
where $\calA(E)$ is an energy-dependent amplitude and $\calQ(E)$ is identified with the Fermi wavevector in the direction of separation between the probes. 

Assuming that each probe couples only to a single site, we find, from \eqref{trans}, that $\calT_{12} \propto |g^0_{12}|^2$. 
The transmission decays monotonically as $1/d_{12}$. Correcting for this geometrical decay yields the constant $\calT \times d$ transmission observed in \figref{T0} for armchair directions. 
The zigzag direction exhibits interference between the $\calQ^+$ and $\calQ^-$ terms entering in \eqref{SPA}.
As seen in the inset of \figref{T0} this leads to both long and short range oscillations. The long range oscillations depend on the Fermi wavelength. 
The short range oscillation on the other hand has a period of three graphene unit cells and is inherent to quantities measured along the zigzag direction and is independent of $E_F$. This oscillation varies on the atomic scale and tends to get cancelled for probes coupling to many sites with different phases. However, the long range oscillations are more robust, particularly for small $E_F$, as the phase is constant over a wider range of sites and should thus be observable even for tips with a larger radius of curvature. The expressions in \eqref{SPA} can also be used to determine the energy-dependent oscillations arising for fixed probes when a gate is applied. Thus the method described here can be easily extended for a spectroscopic mode of a dual-probe system.

\begin{figure}[tb]
\centering
\includegraphics[width=0.90\columnwidth]
{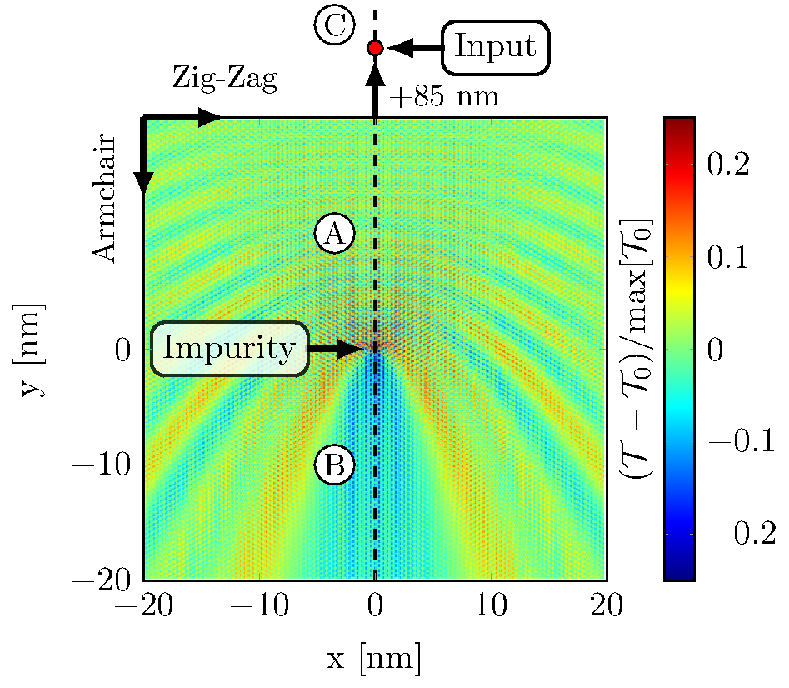}
\caption[]{Relative conductance map for $E_F=0.25 |t|$ around a vacancy at $(0,0)$. The fixed probe (outside the scan area) at $(0,106)$ nm is separated from the impurity along the armchair direction. 
} \label{fig:scatdTac}
\end{figure}

{\it Single Vacancy.}
The GF for a graphene system with a perturbation can be calculated using the Dyson equation,
\bal
G_{ij} = g^0_{ij} + \sum_{nm} g^0_{in} V_{nm} G_{mj},
\eal
where $V_{nm}$ is the perturbation matrix element between site $n$ and $m$.  
In principle any local perturbation can be included using this technique, and accurate parameterization for defects can be determined by comparison with density functional theory calculations \cite{Ribeiro2009,Dubois2010}. The same approach is used throughout to include hopping terms between the probes and device region.

\figref{scatdTac} shows the relative change in transmission from the pristine lattice case when a single vacancy is introduced at the origin. The vacancy and fixed probe are separated along the armchair direction and the scanning probe measures conductance fluctuations in the region around the vacancy. Quantum interference effects are clearly visible in \figref{scatdTac}. The map for a zigzag separation of fixed probe and vacancy (not shown) looks qualitatively similar. To describe the oscillations we again turn to the SPA expression for the GF. The solution of the Dyson equation for a vacancy is $G_{ij}=g^{0}_{ij} + g^{0}_{i0}t_{00} g^{0}_{0j} $, where $t_{00}=-1/g^{0}_{00}$ is the $t$-matrix element of site $0$ when $V_{00} \rightarrow \infty$. 
Analytic solutions can be found for the scanning probe path shown by the dashed line in \figref{scatdTac}. 
We observe oscillations in region A, where the scanning probe is between the fixed probe and vacancy such that $d_{12} =d_{10}-d_{20}$ (see \figref{sketchT0}). From \eqref{SPA} we find  $\Delta \calT \propto \calR e\big[ \calA t_{00} \exp\big(2\ii \calQ d_{20}\big)/\sqrt{d_{10}d_{20}d_{12}} \big] $, which exhibits $2\calQ$ oscillations. When the scanning probe is not between the fixed probe and vacancy no oscillations occur. In region B the transmission is decreased due to scattering, whereas in region C the transmission is either enhanced or decreased depending on the phase difference between the emitted and backscattered waves. 

\begin{figure}[tb]
\centering
\includegraphics[width=0.90\columnwidth]
{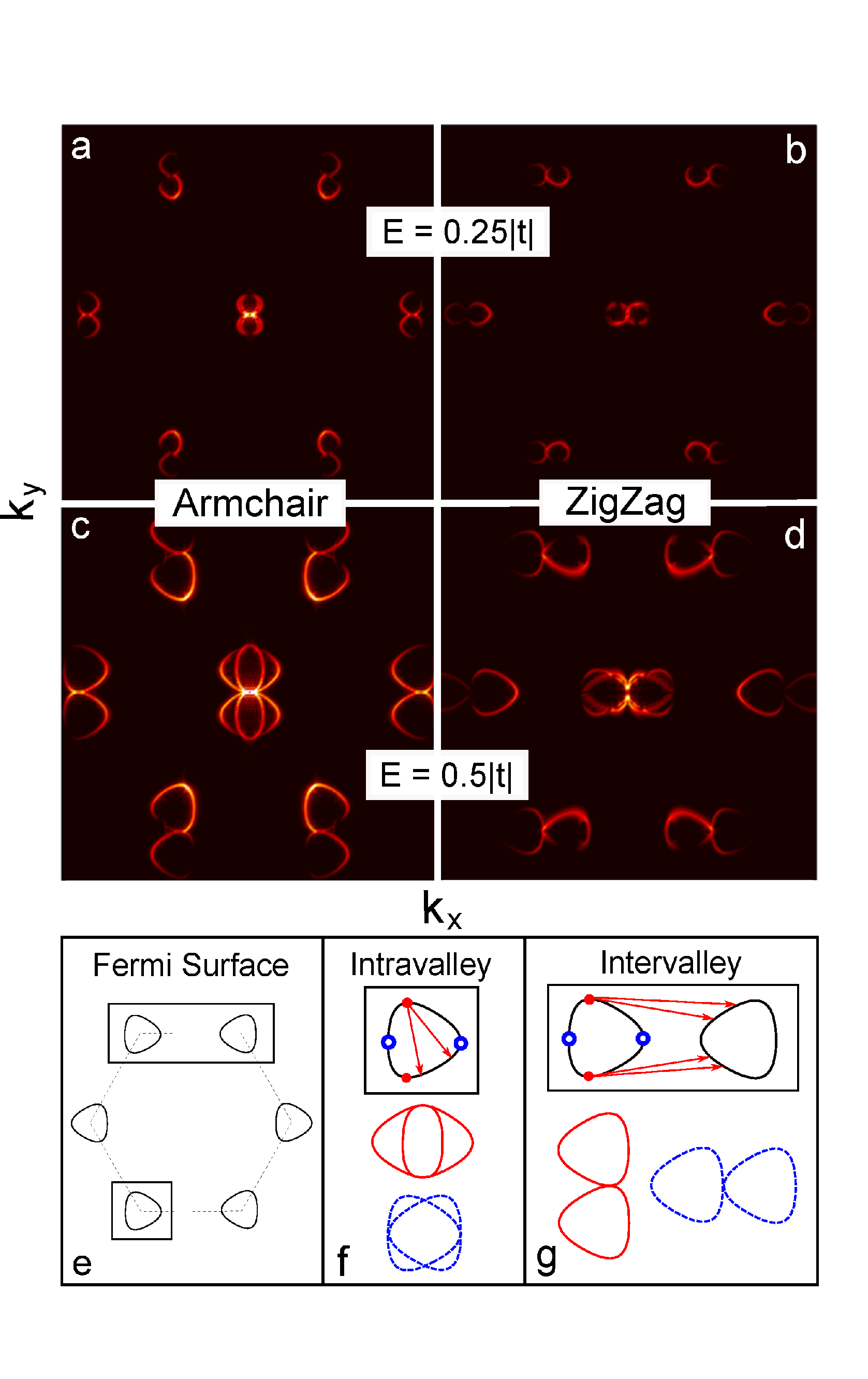}
\caption[]{Fourier transform of the real-space map of $\Delta \calT$ for a single vacancy separated from the fixed probe along the armchair ((a) and (c)), and zigzag ((b) and (d)) directions. Energy is in the linear regime ($E=0.25|t|$) in (a) and (b), and beyond the linear regime ($E=0.5|t|$) in (c) and (d). (e) The Fermi surface of graphene beyond the linear regime. (f)-(g) Scattering diagrams for intra- and intervalley scattering.} \label{fig:FT}
\end{figure}

This simple analytical picture allows us to interpret the oscillations as interferences between an incoming plane wave and the backscattered wave, analogous to optical interference effects. To analyze the pattern we consider the Fourier transform (FT) of the conductance map. This approach is generally applicable to scanning images and is not limited to the graphene example. Similar procedures are often employed in the analysis of conventional STM measurements \cite{Mallet2007,Rutter2007,Deshpande2009}. 
\figref{FT} shows the FTs of $\Delta \calT$ for the single vacancy at different energies and positions of the fixed probe relative to a vacancy at the origin, with Panel (a) corresponding to \figref{scatdTac}. 


For the incoming wave along the $-y$ (armchair) direction the double-ring patterns in \figref{FT} are the result of scattering from the top and bottom of the Fermi surface where the $k$ vectors are along the $y$ direction (indicated by red dots in Figs. \ref{fig:FT}f and \ref{fig:FT}g), to all other points (indicated with arrows) on either the same Fermi surface (intravalley, \figref{FT}f) or that of the opposite valley (intervalley, \figref{FT}g). 
The intravalley scattering produces the short wavevector features present at the center of the FT (and at all reciprocal lattice vectors), while the intervalley scattering yields the larger wavevector features at the $K$ and $K'$-points. Figs. \ref{fig:FT}a and \ref{fig:FT}b correspond to an energy in the linear dispersion regime whereas \ref{fig:FT}c and \ref{fig:FT}d show an energy with trigonal warping, thus leading to the FT signatures sketched by the diagrams of \figref{FT}f-g.

Additional fine structure is seen in \figref{FT} due to deviations from the ideal picture of a plane incoming wave. Allowing a broader range of incoming $k$-vectors increases the part of the Fermi surface which can act as an initial state. This effect is more pronounced for incoming waves along the zigzag direction where even a small broadening of the incoming $k$-vector allows a larger part of the Fermi surface to act as an initial state. 
Similar calculations performed for a Gaussian shaped charge distribution, modelling a trapped charge, find that the FT scattering fingerprint is qualitatively similar to that of the single vacancy. This is in contrast to single-probe LDOS measurements, where the intervalley scattering fingerprint vanishes for extended defects\cite{Wakabayashi2007,Bergvall2013}. 

\begin{figure}[tb]
\centering
\includegraphics[width=0.99\columnwidth]
{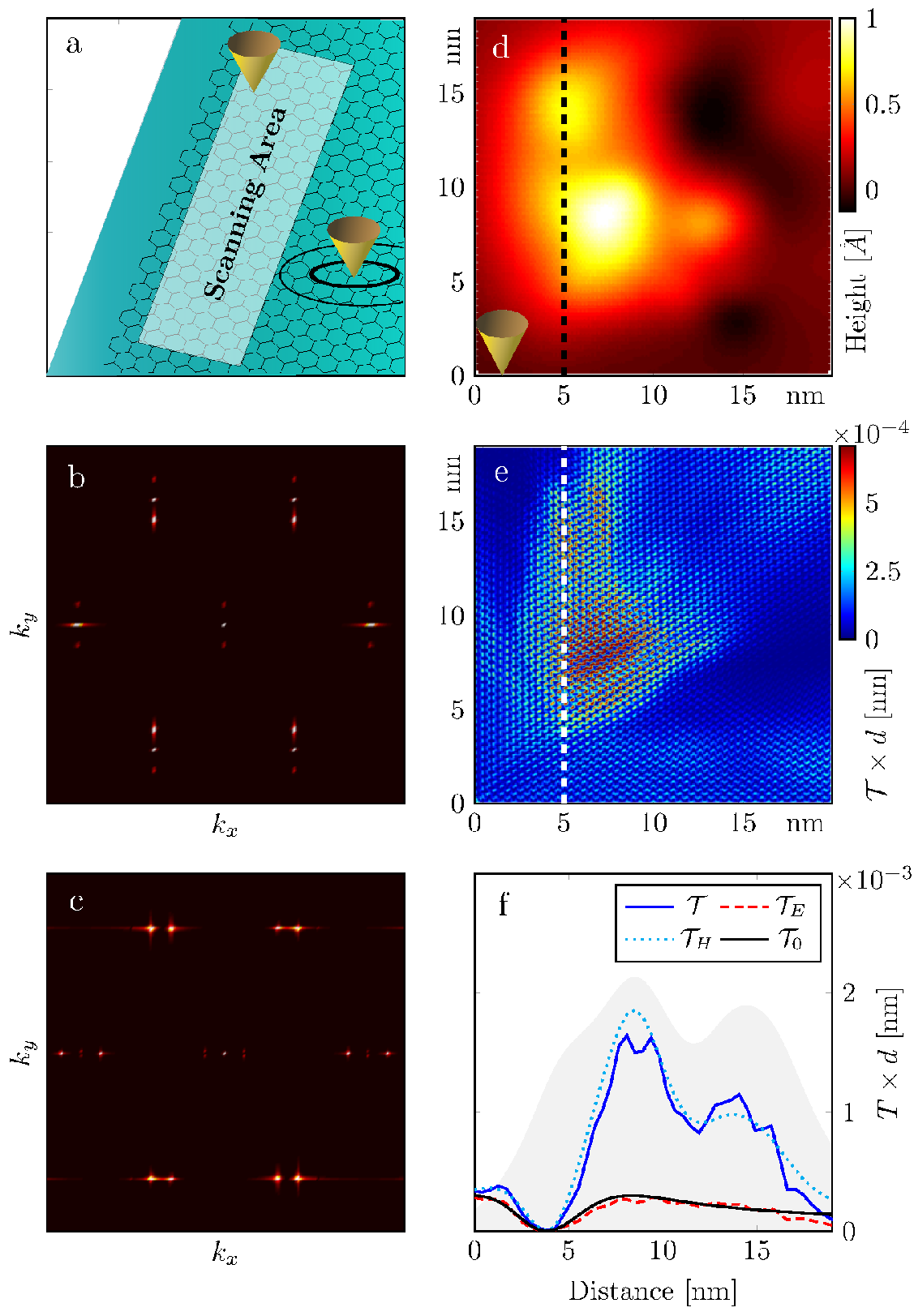}

\caption[]{\small (a) Dual-probe calculation on an edged graphene system. 
 (b)-(c) Fourier transform of the conductance map near a zigzag edge (b) and an armchair edge (c) for $E_F=0.25|t|$. 
 (d) Height profile of a non-planar graphene sheet.  
 (e) Conductance map for the non-planar surface from (d) with the stationary probe at (0,0).
 (f) Transmissions along the dashed line in (b),(e). See main text for further description.
 } \label{fig:FT_edge}
\end{figure}

{\it Other Geometries.}
-- We now consider two examples of more complicated defects: (i) A graphene sheet with an edge (Fig \ref{fig:FT_edge}a), and (ii) a non-planar sheet with an irregular height profile (Fig \ref{fig:FT_edge}d).

In \figref{FT_edge}a, we consider a semi-infinite graphene sheet with a pristine zigzag or armchair edge\footnote{The GF including an armchair edge, is calculated from the pristine GF with the method of images, as described in Ref. \cite{Duffy2013}. For the zigzag edge a direct inversion scheme is used.}. The incoming wave for the armchair edge is along the zigzag direction, and vice versa. The conductance maps (not shown) reveal oscillations away from the edge arising from the interference between incoming and backscattered waves. In contrast to the single vacancy case, not all scattering angles are available and the double-ring features in the FTs reduce to points indicating the direction of propagation (zigzag for armchair edge and vice versa) as shown in Figs. \ref{fig:FT_edge}b and \ref{fig:FT_edge}c. The only qualitative difference is the direction of the incoming wave and hence the direction of the scattering fingerprint in the FT. This is in sharp contrast to single-probe STM measurements, where the zigzag edge does not show an intervalley signal\cite{Casiraghi2009}. The dual-probe setup thus opens the possibility of characterizing an edge by its interference pattern as both edges are equally visible with different signatures.

A non-planar height profile, as in \figref{FT_edge}d, affects the dual-probe measurement in two ways\footnote{Random smooth height fluctuations with amplitudes in accordance to \cite{Fasolino2007}. The electronic effects can be accounted for by varying the hopping parameters according to \cite{Pereira2009} as $t=t_0\ee^{-3.37 (a/a_0-1)}$, where $a$ is the new bond length and $a_0=1.42 \AA$.}. 
The underlying electronic properties of the system are altered by the varying bond lengths throughout the sample and secondly, the tip-sample coupling is affected by their now spatially-varying separation. The conductance map in \figref{FT_edge}e takes both of these effects into account. The signal enhancement for regions where the tip and sample are nearest suggest that it is the tip-sample separation dependent contribution which dominates. This is confirmed in \figref{FT_edge}f where we calculate the full transmission (blue) along the cross section shown by the white dashed line in \figref{FT_edge}e, with the shaded region showing the height profile along this path. In addition, we show the transmissions including the electronic contribution only, $\calT_E$, (dashed red, calculated by mapping the changed electronic structure onto a flat surface) and the height contribution only, $\calT_H$ (dotted green, calculated by varying the tip-sample separation but leaving the sample electronic structure unchanged). 
We note that $\calT_H$ is a good match to the full calculation, whereas $\calT_E$ only slightly deviates from the pristine $T_0$ (black) curve.
However, the height fluctuations considered here are not large enough to give rise to pseudomagnetic field effects like the ones considered in Ref. \cite{Juan2011}. In such cases the behaviour of $\calT_E$ may provide an ideal framework to determine the effects of pseudomagnetic fields on the transport properties.

{\it Conclusion.}
-- The dual-probe setup offers new flexibility to study directional transport effects in nanosystems beyond the reach for a single STM probe experiment. Using graphene as a case study, anisotropic effects in the pristine material and quantum interferences around defects have been treated. The methodology developed is general and easily applicable to other materials. While the focus of this work has been on the scanning mode to reveal topographic details of the sample, an extension to the case of fixed probes and a variable gate gathering spectroscopic data is straightforward. This may be particularly useful when examining non-planar systems, where the variations due to tip-sample separation may outweigh contributions arising from the actual electronic properties of the system.

\textbf{Acknowledgements}
The Center for Nanostructured Graphene (CNG) is sponsored by the Danish Research Foundation, Project DNRF58.


\begin{thebibliography}{57}%
\makeatletter
\providecommand \@ifxundefined [1]{%
 \@ifx{#1\undefined}
}%
\providecommand \@ifnum [1]{%
 \ifnum #1\expandafter \@firstoftwo
 \else \expandafter \@secondoftwo
 \fi
}%
\providecommand \@ifx [1]{%
 \ifx #1\expandafter \@firstoftwo
 \else \expandafter \@secondoftwo
 \fi
}%
\providecommand \natexlab [1]{#1}%
\providecommand \enquote  [1]{``#1''}%
\providecommand \bibnamefont  [1]{#1}%
\providecommand \bibfnamefont [1]{#1}%
\providecommand \citenamefont [1]{#1}%
\providecommand \href@noop [0]{\@secondoftwo}%
\providecommand \href [0]{\begingroup \@sanitize@url \@href}%
\providecommand \@href[1]{\@@startlink{#1}\@@href}%
\providecommand \@@href[1]{\endgroup#1\@@endlink}%
\providecommand \@sanitize@url [0]{\catcode `\\12\catcode `\$12\catcode
  `\&12\catcode `\#12\catcode `\^12\catcode `\_12\catcode `\%12\relax}%
\providecommand \@@startlink[1]{}%
\providecommand \@@endlink[0]{}%
\providecommand \url  [0]{\begingroup\@sanitize@url \@url }%
\providecommand \@url [1]{\endgroup\@href {#1}{\urlprefix }}%
\providecommand \urlprefix  [0]{URL }%
\providecommand \Eprint [0]{\href }%
\providecommand \doibase [0]{http://dx.doi.org/}%
\providecommand \selectlanguage [0]{\@gobble}%
\providecommand \bibinfo  [0]{\@secondoftwo}%
\providecommand \bibfield  [0]{\@secondoftwo}%
\providecommand \translation [1]{[#1]}%
\providecommand \BibitemOpen [0]{}%
\providecommand \bibitemStop [0]{}%
\providecommand \bibitemNoStop [0]{.\EOS\space}%
\providecommand \EOS [0]{\spacefactor3000\relax}%
\providecommand \BibitemShut  [1]{\csname bibitem#1\endcsname}%
\let\auto@bib@innerbib\@empty
\bibitem [{\citenamefont {{Castro Neto}}\ \emph {et~al.}(2009)\citenamefont
  {{Castro Neto}}, \citenamefont {Peres}, \citenamefont {Novoselov},\ and\
  \citenamefont {Geim}}]{CastroNeto2009}%
  \BibitemOpen
  \bibfield  {author} {\bibinfo {author} {\bibfnamefont {A.~H.}\ \bibnamefont
  {{Castro Neto}}}, \bibinfo {author} {\bibfnamefont {N.~M.~R.}\ \bibnamefont
  {Peres}}, \bibinfo {author} {\bibfnamefont {K.~S.}\ \bibnamefont
  {Novoselov}}, \ and\ \bibinfo {author} {\bibfnamefont {A.~K.}\ \bibnamefont
  {Geim}},\ }\href {\doibase 10.1103/RevModPhys.81.109} {\bibfield  {journal}
  {\bibinfo  {journal} {Reviews of Modern Physics}\ }\textbf {\bibinfo {volume}
  {81}},\ \bibinfo {pages} {109} (\bibinfo {year} {2009})}\BibitemShut
  {NoStop}%
\bibitem [{\citenamefont {Chen}\ \emph {et~al.}(2008)\citenamefont {Chen},
  \citenamefont {Jang}, \citenamefont {Xiao}, \citenamefont {Ishigami},\ and\
  \citenamefont {Fuhrer}}]{Chen2008}%
  \BibitemOpen
  \bibfield  {author} {\bibinfo {author} {\bibfnamefont {J.-H.}\ \bibnamefont
  {Chen}}, \bibinfo {author} {\bibfnamefont {C.}~\bibnamefont {Jang}}, \bibinfo
  {author} {\bibfnamefont {S.}~\bibnamefont {Xiao}}, \bibinfo {author}
  {\bibfnamefont {M.}~\bibnamefont {Ishigami}}, \ and\ \bibinfo {author}
  {\bibfnamefont {M.~S.}\ \bibnamefont {Fuhrer}},\ }\href {\doibase
  10.1038/nnano.2008.58} {\bibfield  {journal} {\bibinfo  {journal} {Nature
  nanotechnology}\ }\textbf {\bibinfo {volume} {3}},\ \bibinfo {pages} {206}
  (\bibinfo {year} {2008})}\BibitemShut {NoStop}%
\bibitem [{\citenamefont {Hwang}\ \emph {et~al.}(2007)\citenamefont {Hwang},
  \citenamefont {Adam},\ and\ \citenamefont {Das~Sarma}}]{Hwang2007}%
  \BibitemOpen
  \bibfield  {author} {\bibinfo {author} {\bibfnamefont {E.~H.}\ \bibnamefont
  {Hwang}}, \bibinfo {author} {\bibfnamefont {S.}~\bibnamefont {Adam}}, \ and\
  \bibinfo {author} {\bibfnamefont {S.}~\bibnamefont {Das~Sarma}},\ }\href
  {\doibase 10.1103/PhysRevLett.98.186806} {\bibfield  {journal} {\bibinfo
  {journal} {Physical Review Letters}\ }\textbf {\bibinfo {volume} {98}},\
  \bibinfo {pages} {186806} (\bibinfo {year} {2007})}\BibitemShut {NoStop}%
\bibitem [{\citenamefont {Dean}\ \emph {et~al.}(2010)\citenamefont {Dean},
  \citenamefont {Young}, \citenamefont {Meric}, \citenamefont {Lee},
  \citenamefont {Wang}, \citenamefont {Sorgenfrei}, \citenamefont {Watanabe},
  \citenamefont {Taniguchi}, \citenamefont {Kim}, \citenamefont {Shepard},\
  and\ \citenamefont {Hone}}]{Dean2010}%
  \BibitemOpen
  \bibfield  {author} {\bibinfo {author} {\bibfnamefont {C.~R.}\ \bibnamefont
  {Dean}}, \bibinfo {author} {\bibfnamefont {A.~F.}\ \bibnamefont {Young}},
  \bibinfo {author} {\bibfnamefont {I.}~\bibnamefont {Meric}}, \bibinfo
  {author} {\bibfnamefont {C.}~\bibnamefont {Lee}}, \bibinfo {author}
  {\bibfnamefont {L.}~\bibnamefont {Wang}}, \bibinfo {author} {\bibfnamefont
  {S.}~\bibnamefont {Sorgenfrei}}, \bibinfo {author} {\bibfnamefont
  {K.}~\bibnamefont {Watanabe}}, \bibinfo {author} {\bibfnamefont
  {T.}~\bibnamefont {Taniguchi}}, \bibinfo {author} {\bibfnamefont
  {P.}~\bibnamefont {Kim}}, \bibinfo {author} {\bibfnamefont {K.~L.}\
  \bibnamefont {Shepard}}, \ and\ \bibinfo {author} {\bibfnamefont
  {J.}~\bibnamefont {Hone}},\ }\href {\doibase 10.1038/nnano.2010.172}
  {\bibfield  {journal} {\bibinfo  {journal} {Nature Nanotechnology}\ }\textbf
  {\bibinfo {volume} {5}},\ \bibinfo {pages} {722} (\bibinfo {year}
  {2010})}\BibitemShut {NoStop}%
\bibitem [{\citenamefont {Schedin}\ \emph {et~al.}(2007)\citenamefont
  {Schedin}, \citenamefont {Geim}, \citenamefont {Morozov}, \citenamefont
  {Hill}, \citenamefont {Blake}, \citenamefont {Katsnelson},\ and\
  \citenamefont {Novoselov}}]{Schedin2007}%
  \BibitemOpen
  \bibfield  {author} {\bibinfo {author} {\bibfnamefont {F.}~\bibnamefont
  {Schedin}}, \bibinfo {author} {\bibfnamefont {A.~K.}\ \bibnamefont {Geim}},
  \bibinfo {author} {\bibfnamefont {S.~V.}\ \bibnamefont {Morozov}}, \bibinfo
  {author} {\bibfnamefont {E.~W.}\ \bibnamefont {Hill}}, \bibinfo {author}
  {\bibfnamefont {P.}~\bibnamefont {Blake}}, \bibinfo {author} {\bibfnamefont
  {M.~I.}\ \bibnamefont {Katsnelson}}, \ and\ \bibinfo {author} {\bibfnamefont
  {K.~S.}\ \bibnamefont {Novoselov}},\ }\href {\doibase 10.1038/nmat1967}
  {\bibfield  {journal} {\bibinfo  {journal} {Nature Materials}\ }\textbf
  {\bibinfo {volume} {6}},\ \bibinfo {pages} {652} (\bibinfo {year}
  {2007})}\BibitemShut {NoStop}%
\bibitem [{\citenamefont {Brar}\ \emph {et~al.}(2010)\citenamefont {Brar},
  \citenamefont {Decker}, \citenamefont {Solowan}, \citenamefont {Wang},
  \citenamefont {Maserati}, \citenamefont {Chan}, \citenamefont {Lee},
  \citenamefont {Girit}, \citenamefont {Zettl}, \citenamefont {Louie},
  \citenamefont {Cohen},\ and\ \citenamefont {Crommie}}]{Brar2010}%
  \BibitemOpen
  \bibfield  {author} {\bibinfo {author} {\bibfnamefont {V.~W.}\ \bibnamefont
  {Brar}}, \bibinfo {author} {\bibfnamefont {R.}~\bibnamefont {Decker}},
  \bibinfo {author} {\bibfnamefont {H.-M.}\ \bibnamefont {Solowan}}, \bibinfo
  {author} {\bibfnamefont {Y.}~\bibnamefont {Wang}}, \bibinfo {author}
  {\bibfnamefont {L.}~\bibnamefont {Maserati}}, \bibinfo {author}
  {\bibfnamefont {K.~T.}\ \bibnamefont {Chan}}, \bibinfo {author}
  {\bibfnamefont {H.}~\bibnamefont {Lee}}, \bibinfo {author} {\bibfnamefont
  {c.~O.}\ \bibnamefont {Girit}}, \bibinfo {author} {\bibfnamefont
  {A.}~\bibnamefont {Zettl}}, \bibinfo {author} {\bibfnamefont {S.~G.}\
  \bibnamefont {Louie}}, \bibinfo {author} {\bibfnamefont {M.~L.}\ \bibnamefont
  {Cohen}}, \ and\ \bibinfo {author} {\bibfnamefont {M.~F.}\ \bibnamefont
  {Crommie}},\ }\href {\doibase 10.1038/nphys1807} {\bibfield  {journal}
  {\bibinfo  {journal} {Nature Physics}\ }\textbf {\bibinfo {volume} {7}},\
  \bibinfo {pages} {43} (\bibinfo {year} {2010})}\BibitemShut {NoStop}%
\bibitem [{\citenamefont {Peres}\ \emph {et~al.}(2006)\citenamefont {Peres},
  \citenamefont {Guinea},\ and\ \citenamefont {{Castro Neto}}}]{Peres2006}%
  \BibitemOpen
  \bibfield  {author} {\bibinfo {author} {\bibfnamefont {N.~M.~R.}\
  \bibnamefont {Peres}}, \bibinfo {author} {\bibfnamefont {F.}~\bibnamefont
  {Guinea}}, \ and\ \bibinfo {author} {\bibfnamefont {A.~H.}\ \bibnamefont
  {{Castro Neto}}},\ }\href {\doibase 10.1103/PhysRevB.73.125411} {\bibfield
  {journal} {\bibinfo  {journal} {Physical Review B}\ }\textbf {\bibinfo
  {volume} {73}},\ \bibinfo {pages} {125411} (\bibinfo {year}
  {2006})}\BibitemShut {NoStop}%
\bibitem [{\citenamefont {Peres}(2010)}]{Peres2010}%
  \BibitemOpen
  \bibfield  {author} {\bibinfo {author} {\bibfnamefont {N.~M.~R.}\
  \bibnamefont {Peres}},\ }\href {\doibase 10.1103/RevModPhys.82.2673}
  {\bibfield  {journal} {\bibinfo  {journal} {Reviews of Modern Physics}\
  }\textbf {\bibinfo {volume} {82}},\ \bibinfo {pages} {2673} (\bibinfo {year}
  {2010})}\BibitemShut {NoStop}%
\bibitem [{\citenamefont {Bena}(2008)}]{Bena2008}%
  \BibitemOpen
  \bibfield  {author} {\bibinfo {author} {\bibfnamefont {C.}~\bibnamefont
  {Bena}},\ }\href {\doibase 10.1103/PhysRevLett.100.076601} {\bibfield
  {journal} {\bibinfo  {journal} {Physical Review Letters}\ }\textbf {\bibinfo
  {volume} {100}},\ \bibinfo {pages} {076601} (\bibinfo {year}
  {2008})}\BibitemShut {NoStop}%
\bibitem [{\citenamefont {Simon}\ \emph {et~al.}(2009)\citenamefont {Simon},
  \citenamefont {Bena}, \citenamefont {Vonau}, \citenamefont {Aubel},
  \citenamefont {Nasrallah}, \citenamefont {Habar},\ and\ \citenamefont
  {Peruchetti}}]{Simon2009}%
  \BibitemOpen
  \bibfield  {author} {\bibinfo {author} {\bibfnamefont {L.}~\bibnamefont
  {Simon}}, \bibinfo {author} {\bibfnamefont {C.}~\bibnamefont {Bena}},
  \bibinfo {author} {\bibfnamefont {F.}~\bibnamefont {Vonau}}, \bibinfo
  {author} {\bibfnamefont {D.}~\bibnamefont {Aubel}}, \bibinfo {author}
  {\bibfnamefont {H.}~\bibnamefont {Nasrallah}}, \bibinfo {author}
  {\bibfnamefont {M.}~\bibnamefont {Habar}}, \ and\ \bibinfo {author}
  {\bibfnamefont {J.~C.}\ \bibnamefont {Peruchetti}},\ }\href {\doibase
  10.1140/epjb/e2009-00142-3} {\bibfield  {journal} {\bibinfo  {journal} {The
  European Physical Journal B}\ }\textbf {\bibinfo {volume} {69}},\ \bibinfo
  {pages} {351} (\bibinfo {year} {2009})}\BibitemShut {NoStop}%
\bibitem [{\citenamefont {Amara}\ \emph {et~al.}(2007)\citenamefont {Amara},
  \citenamefont {Latil}, \citenamefont {Meunier}, \citenamefont {Lambin},\ and\
  \citenamefont {Charlier}}]{Amara2007}%
  \BibitemOpen
  \bibfield  {author} {\bibinfo {author} {\bibfnamefont {H.}~\bibnamefont
  {Amara}}, \bibinfo {author} {\bibfnamefont {S.}~\bibnamefont {Latil}},
  \bibinfo {author} {\bibfnamefont {V.}~\bibnamefont {Meunier}}, \bibinfo
  {author} {\bibfnamefont {P.}~\bibnamefont {Lambin}}, \ and\ \bibinfo {author}
  {\bibfnamefont {J.-C.}\ \bibnamefont {Charlier}},\ }\href {\doibase
  10.1103/PhysRevB.76.115423} {\bibfield  {journal} {\bibinfo  {journal}
  {Physical Review B}\ }\textbf {\bibinfo {volume} {76}},\ \bibinfo {pages}
  {115423} (\bibinfo {year} {2007})}\BibitemShut {NoStop}%
\bibitem [{\citenamefont {Mallet}\ \emph {et~al.}(2007)\citenamefont {Mallet},
  \citenamefont {Varchon}, \citenamefont {Naud}, \citenamefont {Magaud},
  \citenamefont {Berger},\ and\ \citenamefont {Veuillen}}]{Mallet2007}%
  \BibitemOpen
  \bibfield  {author} {\bibinfo {author} {\bibfnamefont {P.}~\bibnamefont
  {Mallet}}, \bibinfo {author} {\bibfnamefont {F.}~\bibnamefont {Varchon}},
  \bibinfo {author} {\bibfnamefont {C.}~\bibnamefont {Naud}}, \bibinfo {author}
  {\bibfnamefont {L.}~\bibnamefont {Magaud}}, \bibinfo {author} {\bibfnamefont
  {C.}~\bibnamefont {Berger}}, \ and\ \bibinfo {author} {\bibfnamefont {J.-Y.}\
  \bibnamefont {Veuillen}},\ }\href {\doibase 10.1103/PhysRevB.76.041403}
  {\bibfield  {journal} {\bibinfo  {journal} {Physical Review B}\ }\textbf
  {\bibinfo {volume} {76}},\ \bibinfo {pages} {041403} (\bibinfo {year}
  {2007})}\BibitemShut {NoStop}%
\bibitem [{\citenamefont {Rutter}\ \emph {et~al.}(2007)\citenamefont {Rutter},
  \citenamefont {Crain}, \citenamefont {Guisinger}, \citenamefont {Li},
  \citenamefont {First},\ and\ \citenamefont {Stroscio}}]{Rutter2007}%
  \BibitemOpen
  \bibfield  {author} {\bibinfo {author} {\bibfnamefont {G.~M.}\ \bibnamefont
  {Rutter}}, \bibinfo {author} {\bibfnamefont {J.~N.}\ \bibnamefont {Crain}},
  \bibinfo {author} {\bibfnamefont {N.~P.}\ \bibnamefont {Guisinger}}, \bibinfo
  {author} {\bibfnamefont {T.}~\bibnamefont {Li}}, \bibinfo {author}
  {\bibfnamefont {P.~N.}\ \bibnamefont {First}}, \ and\ \bibinfo {author}
  {\bibfnamefont {J.~A.}\ \bibnamefont {Stroscio}},\ }\href {\doibase
  10.1126/science.1142882} {\bibfield  {journal} {\bibinfo  {journal}
  {Science}\ }\textbf {\bibinfo {volume} {317}},\ \bibinfo {pages} {219}
  (\bibinfo {year} {2007})}\BibitemShut {NoStop}%
\bibitem [{\citenamefont {Peres}\ \emph {et~al.}(2009)\citenamefont {Peres},
  \citenamefont {Yang},\ and\ \citenamefont {Tsai}}]{Peres2009}%
  \BibitemOpen
  \bibfield  {author} {\bibinfo {author} {\bibfnamefont {N.~M.~R.}\
  \bibnamefont {Peres}}, \bibinfo {author} {\bibfnamefont {L.}~\bibnamefont
  {Yang}}, \ and\ \bibinfo {author} {\bibfnamefont {S.-W.}\ \bibnamefont
  {Tsai}},\ }\href {\doibase 10.1088/1367-2630/11/9/095007} {\bibfield
  {journal} {\bibinfo  {journal} {New Journal of Physics}\ }\textbf {\bibinfo
  {volume} {11}},\ \bibinfo {pages} {095007} (\bibinfo {year}
  {2009})}\BibitemShut {NoStop}%
\bibitem [{\citenamefont {Cheianov}\ and\ \citenamefont
  {Fal'ko}(2006)}]{Cheianov2006}%
  \BibitemOpen
  \bibfield  {author} {\bibinfo {author} {\bibfnamefont {V.~V.}\ \bibnamefont
  {Cheianov}}\ and\ \bibinfo {author} {\bibfnamefont {V.~I.}\ \bibnamefont
  {Fal'ko}},\ }\href {\doibase 10.1103/PhysRevLett.97.226801} {\bibfield
  {journal} {\bibinfo  {journal} {Physical Review Letters}\ }\textbf {\bibinfo
  {volume} {97}},\ \bibinfo {pages} {226801} (\bibinfo {year}
  {2006})}\BibitemShut {NoStop}%
\bibitem [{\citenamefont {B\'{a}csi}\ and\ \citenamefont
  {Virosztek}(2010)}]{Bacsi2010}%
  \BibitemOpen
  \bibfield  {author} {\bibinfo {author} {\bibfnamefont {A.}~\bibnamefont
  {B\'{a}csi}}\ and\ \bibinfo {author} {\bibfnamefont {A.}~\bibnamefont
  {Virosztek}},\ }\href {\doibase 10.1103/PhysRevB.82.193405} {\bibfield
  {journal} {\bibinfo  {journal} {Physical Review B}\ }\textbf {\bibinfo
  {volume} {82}},\ \bibinfo {pages} {193405} (\bibinfo {year}
  {2010})}\BibitemShut {NoStop}%
\bibitem [{\citenamefont {Xue}\ \emph {et~al.}(2012)\citenamefont {Xue},
  \citenamefont {Sanchez-Yamagishi}, \citenamefont {Watanabe}, \citenamefont
  {Taniguchi}, \citenamefont {Jarillo-Herrero},\ and\ \citenamefont
  {LeRoy}}]{Xue2012}%
  \BibitemOpen
  \bibfield  {author} {\bibinfo {author} {\bibfnamefont {J.}~\bibnamefont
  {Xue}}, \bibinfo {author} {\bibfnamefont {J.}~\bibnamefont
  {Sanchez-Yamagishi}}, \bibinfo {author} {\bibfnamefont {K.}~\bibnamefont
  {Watanabe}}, \bibinfo {author} {\bibfnamefont {T.}~\bibnamefont {Taniguchi}},
  \bibinfo {author} {\bibfnamefont {P.}~\bibnamefont {Jarillo-Herrero}}, \ and\
  \bibinfo {author} {\bibfnamefont {B.~J.}\ \bibnamefont {LeRoy}},\ }\href
  {\doibase 10.1103/PhysRevLett.108.016801} {\bibfield  {journal} {\bibinfo
  {journal} {Physical Review Letters}\ }\textbf {\bibinfo {volume} {108}},\
  \bibinfo {pages} {016801} (\bibinfo {year} {2012})}\BibitemShut {NoStop}%
\bibitem [{\citenamefont {Yang}\ \emph {et~al.}(2010)\citenamefont {Yang},
  \citenamefont {Mayne}, \citenamefont {Boucherit}, \citenamefont {Comtet},
  \citenamefont {Dujardin},\ and\ \citenamefont {Kuk}}]{Yang2010}%
  \BibitemOpen
  \bibfield  {author} {\bibinfo {author} {\bibfnamefont {H.}~\bibnamefont
  {Yang}}, \bibinfo {author} {\bibfnamefont {A.~J.}\ \bibnamefont {Mayne}},
  \bibinfo {author} {\bibfnamefont {M.}~\bibnamefont {Boucherit}}, \bibinfo
  {author} {\bibfnamefont {G.}~\bibnamefont {Comtet}}, \bibinfo {author}
  {\bibfnamefont {G.}~\bibnamefont {Dujardin}}, \ and\ \bibinfo {author}
  {\bibfnamefont {Y.}~\bibnamefont {Kuk}},\ }\href {\doibase 10.1021/nl9038778}
  {\bibfield  {journal} {\bibinfo  {journal} {Nano Letters}\ }\textbf {\bibinfo
  {volume} {10}},\ \bibinfo {pages} {943} (\bibinfo {year} {2010})}\BibitemShut
  {NoStop}%
\bibitem [{\citenamefont {Park}\ \emph {et~al.}(2011)\citenamefont {Park},
  \citenamefont {Yang}, \citenamefont {Mayne}, \citenamefont {Dujardin},
  \citenamefont {Seo}, \citenamefont {Kuk}, \citenamefont {Ihm},\ and\
  \citenamefont {Kim}}]{Park2011}%
  \BibitemOpen
  \bibfield  {author} {\bibinfo {author} {\bibfnamefont {C.}~\bibnamefont
  {Park}}, \bibinfo {author} {\bibfnamefont {H.}~\bibnamefont {Yang}}, \bibinfo
  {author} {\bibfnamefont {A.~J.}\ \bibnamefont {Mayne}}, \bibinfo {author}
  {\bibfnamefont {G.}~\bibnamefont {Dujardin}}, \bibinfo {author}
  {\bibfnamefont {S.}~\bibnamefont {Seo}}, \bibinfo {author} {\bibfnamefont
  {Y.}~\bibnamefont {Kuk}}, \bibinfo {author} {\bibfnamefont {J.}~\bibnamefont
  {Ihm}}, \ and\ \bibinfo {author} {\bibfnamefont {G.}~\bibnamefont {Kim}},\
  }\href {\doibase 10.1073/pnas.1114548108} {\bibfield  {journal} {\bibinfo
  {journal} {Proceedings of the National Academy of Sciences}\ }\textbf
  {\bibinfo {volume} {108}},\ \bibinfo {pages} {18622} (\bibinfo {year}
  {2011})}\BibitemShut {NoStop}%
\bibitem [{\citenamefont {Barone}\ \emph {et~al.}(2006)\citenamefont {Barone},
  \citenamefont {Hod},\ and\ \citenamefont {Scuseria}}]{Barone2006}%
  \BibitemOpen
  \bibfield  {author} {\bibinfo {author} {\bibfnamefont {V.}~\bibnamefont
  {Barone}}, \bibinfo {author} {\bibfnamefont {O.}~\bibnamefont {Hod}}, \ and\
  \bibinfo {author} {\bibfnamefont {G.~E.}\ \bibnamefont {Scuseria}},\ }\href
  {\doibase 10.1021/nl0617033} {\bibfield  {journal} {\bibinfo  {journal} {Nano
  Letters}\ }\textbf {\bibinfo {volume} {6}},\ \bibinfo {pages} {2748}
  (\bibinfo {year} {2006})}\BibitemShut {NoStop}%
\bibitem [{\citenamefont {Mason}\ \emph {et~al.}(2013)\citenamefont {Mason},
  \citenamefont {Borunda},\ and\ \citenamefont {Heller}}]{Mason2013}%
  \BibitemOpen
  \bibfield  {author} {\bibinfo {author} {\bibfnamefont {D.~J.}\ \bibnamefont
  {Mason}}, \bibinfo {author} {\bibfnamefont {M.~F.}\ \bibnamefont {Borunda}},
  \ and\ \bibinfo {author} {\bibfnamefont {E.~J.}\ \bibnamefont {Heller}},\
  }\href {\doibase 10.1103/PhysRevB.88.165421} {\bibfield  {journal} {\bibinfo
  {journal} {Physical Review B}\ }\textbf {\bibinfo {volume} {88}},\ \bibinfo
  {pages} {165421} (\bibinfo {year} {2013})}\BibitemShut {NoStop}%
\bibitem [{\citenamefont {Bergvall}\ and\ \citenamefont
  {L\"{o}fwander}(2013)}]{Bergvall2013}%
  \BibitemOpen
  \bibfield  {author} {\bibinfo {author} {\bibfnamefont {A.}~\bibnamefont
  {Bergvall}}\ and\ \bibinfo {author} {\bibfnamefont {T.}~\bibnamefont
  {L\"{o}fwander}},\ }\href {\doibase 10.1103/PhysRevB.87.205431} {\bibfield
  {journal} {\bibinfo  {journal} {Physical Review B}\ }\textbf {\bibinfo
  {volume} {87}},\ \bibinfo {pages} {205431} (\bibinfo {year}
  {2013})}\BibitemShut {NoStop}%
\bibitem [{\citenamefont {Deshpande}\ \emph {et~al.}(2009)\citenamefont
  {Deshpande}, \citenamefont {Bao}, \citenamefont {Miao}, \citenamefont {Lau},\
  and\ \citenamefont {LeRoy}}]{Deshpande2009}%
  \BibitemOpen
  \bibfield  {author} {\bibinfo {author} {\bibfnamefont {A.}~\bibnamefont
  {Deshpande}}, \bibinfo {author} {\bibfnamefont {W.}~\bibnamefont {Bao}},
  \bibinfo {author} {\bibfnamefont {F.}~\bibnamefont {Miao}}, \bibinfo {author}
  {\bibfnamefont {C.~N.}\ \bibnamefont {Lau}}, \ and\ \bibinfo {author}
  {\bibfnamefont {B.~J.}\ \bibnamefont {LeRoy}},\ }\href {\doibase
  10.1103/PhysRevB.79.205411} {\bibfield  {journal} {\bibinfo  {journal}
  {Physical Review B}\ }\textbf {\bibinfo {volume} {79}},\ \bibinfo {pages}
  {205411} (\bibinfo {year} {2009})}\BibitemShut {NoStop}%
\bibitem [{\citenamefont {Zhang}\ \emph {et~al.}(2009)\citenamefont {Zhang},
  \citenamefont {Brar}, \citenamefont {Girit}, \citenamefont {Zettl},\ and\
  \citenamefont {Crommie}}]{Zhang2009}%
  \BibitemOpen
  \bibfield  {author} {\bibinfo {author} {\bibfnamefont {Y.}~\bibnamefont
  {Zhang}}, \bibinfo {author} {\bibfnamefont {V.~W.}\ \bibnamefont {Brar}},
  \bibinfo {author} {\bibfnamefont {C.}~\bibnamefont {Girit}}, \bibinfo
  {author} {\bibfnamefont {A.}~\bibnamefont {Zettl}}, \ and\ \bibinfo {author}
  {\bibfnamefont {M.~F.}\ \bibnamefont {Crommie}},\ }\href {\doibase
  10.1038/nphys1365} {\bibfield  {journal} {\bibinfo  {journal} {Nature
  Physics}\ }\textbf {\bibinfo {volume} {5}},\ \bibinfo {pages} {722} (\bibinfo
  {year} {2009})}\BibitemShut {NoStop}%
\bibitem [{\citenamefont {Nakayama}\ \emph {et~al.}(2012)\citenamefont
  {Nakayama}, \citenamefont {Kubo}, \citenamefont {Shingaya}, \citenamefont
  {Higuchi}, \citenamefont {Hasegawa}, \citenamefont {Jiang}, \citenamefont
  {Okuda}, \citenamefont {Kuwahara}, \citenamefont {Takami},\ and\
  \citenamefont {Aono}}]{Nakayama2012}%
  \BibitemOpen
  \bibfield  {author} {\bibinfo {author} {\bibfnamefont {T.}~\bibnamefont
  {Nakayama}}, \bibinfo {author} {\bibfnamefont {O.}~\bibnamefont {Kubo}},
  \bibinfo {author} {\bibfnamefont {Y.}~\bibnamefont {Shingaya}}, \bibinfo
  {author} {\bibfnamefont {S.}~\bibnamefont {Higuchi}}, \bibinfo {author}
  {\bibfnamefont {T.}~\bibnamefont {Hasegawa}}, \bibinfo {author}
  {\bibfnamefont {C.-S.}\ \bibnamefont {Jiang}}, \bibinfo {author}
  {\bibfnamefont {T.}~\bibnamefont {Okuda}}, \bibinfo {author} {\bibfnamefont
  {Y.}~\bibnamefont {Kuwahara}}, \bibinfo {author} {\bibfnamefont
  {K.}~\bibnamefont {Takami}}, \ and\ \bibinfo {author} {\bibfnamefont
  {M.}~\bibnamefont {Aono}},\ }\href {\doibase 10.1002/adma.201200257}
  {\bibfield  {journal} {\bibinfo  {journal} {Advanced materials}\ }\textbf
  {\bibinfo {volume} {24}},\ \bibinfo {pages} {1675} (\bibinfo {year}
  {2012})}\BibitemShut {NoStop}%
\bibitem [{\citenamefont {Qin}\ \emph {et~al.}(2012)\citenamefont {Qin},
  \citenamefont {Kim}, \citenamefont {Wang},\ and\ \citenamefont
  {Li}}]{Qin2012}%
  \BibitemOpen
  \bibfield  {author} {\bibinfo {author} {\bibfnamefont {S.}~\bibnamefont
  {Qin}}, \bibinfo {author} {\bibfnamefont {T.-H.}\ \bibnamefont {Kim}},
  \bibinfo {author} {\bibfnamefont {Z.}~\bibnamefont {Wang}}, \ and\ \bibinfo
  {author} {\bibfnamefont {A.-P.}\ \bibnamefont {Li}},\ }\href {\doibase
  10.1063/1.4727878} {\bibfield  {journal} {\bibinfo  {journal} {Review of
  Scientific Instruments}\ }\textbf {\bibinfo {volume} {83}},\ \bibinfo {pages}
  {063704} (\bibinfo {year} {2012})}\BibitemShut {NoStop}%
\bibitem [{\citenamefont {Cherepanov}\ \emph {et~al.}(2012)\citenamefont
  {Cherepanov}, \citenamefont {Zubkov}, \citenamefont {Junker}, \citenamefont
  {Korte}, \citenamefont {Blab}, \citenamefont {Coenen},\ and\ \citenamefont
  {Voigtl\"{a}nder}}]{Cherepanov2012}%
  \BibitemOpen
  \bibfield  {author} {\bibinfo {author} {\bibfnamefont {V.}~\bibnamefont
  {Cherepanov}}, \bibinfo {author} {\bibfnamefont {E.}~\bibnamefont {Zubkov}},
  \bibinfo {author} {\bibfnamefont {H.}~\bibnamefont {Junker}}, \bibinfo
  {author} {\bibfnamefont {S.}~\bibnamefont {Korte}}, \bibinfo {author}
  {\bibfnamefont {M.}~\bibnamefont {Blab}}, \bibinfo {author} {\bibfnamefont
  {P.}~\bibnamefont {Coenen}}, \ and\ \bibinfo {author} {\bibfnamefont
  {B.}~\bibnamefont {Voigtl\"{a}nder}},\ }\href {\doibase 10.1063/1.3694990}
  {\bibfield  {journal} {\bibinfo  {journal} {Review of scientific
  instruments}\ }\textbf {\bibinfo {volume} {83}},\ \bibinfo {pages} {033707}
  (\bibinfo {year} {2012})}\BibitemShut {NoStop}%
\bibitem [{\citenamefont {Bannani}\ \emph {et~al.}(2008)\citenamefont
  {Bannani}, \citenamefont {Bobisch},\ and\ \citenamefont
  {Moller}}]{Bannani2008}%
  \BibitemOpen
  \bibfield  {author} {\bibinfo {author} {\bibfnamefont {A.}~\bibnamefont
  {Bannani}}, \bibinfo {author} {\bibfnamefont {C.~A.}\ \bibnamefont
  {Bobisch}}, \ and\ \bibinfo {author} {\bibfnamefont {R.}~\bibnamefont
  {Moller}},\ }\href {\doibase 10.1063/1.2968111} {\bibfield  {journal}
  {\bibinfo  {journal} {Review of scientific instruments}\ }\textbf {\bibinfo
  {volume} {79}},\ \bibinfo {pages} {083704} (\bibinfo {year}
  {2008})}\BibitemShut {NoStop}%
\bibitem [{\citenamefont {Ji}\ \emph {et~al.}(2012)\citenamefont {Ji},
  \citenamefont {Hannon}, \citenamefont {Tromp}, \citenamefont {Perebeinos},
  \citenamefont {Tersoff},\ and\ \citenamefont {Ross}}]{Ji2012}%
  \BibitemOpen
  \bibfield  {author} {\bibinfo {author} {\bibfnamefont {S.-H.}\ \bibnamefont
  {Ji}}, \bibinfo {author} {\bibfnamefont {J.~B.}\ \bibnamefont {Hannon}},
  \bibinfo {author} {\bibfnamefont {R.~M.}\ \bibnamefont {Tromp}}, \bibinfo
  {author} {\bibfnamefont {V.}~\bibnamefont {Perebeinos}}, \bibinfo {author}
  {\bibfnamefont {J.}~\bibnamefont {Tersoff}}, \ and\ \bibinfo {author}
  {\bibfnamefont {F.~M.}\ \bibnamefont {Ross}},\ }\href {\doibase
  10.1038/nmat3170} {\bibfield  {journal} {\bibinfo  {journal} {Nature
  materials}\ }\textbf {\bibinfo {volume} {11}},\ \bibinfo {pages} {114}
  (\bibinfo {year} {2012})}\BibitemShut {NoStop}%
\bibitem [{\citenamefont {Sutter}\ \emph {et~al.}(2008)\citenamefont {Sutter},
  \citenamefont {Flege},\ and\ \citenamefont {Sutter}}]{Sutter2008}%
  \BibitemOpen
  \bibfield  {author} {\bibinfo {author} {\bibfnamefont {P.~W.}\ \bibnamefont
  {Sutter}}, \bibinfo {author} {\bibfnamefont {J.-I.}\ \bibnamefont {Flege}}, \
  and\ \bibinfo {author} {\bibfnamefont {E.~A.}\ \bibnamefont {Sutter}},\
  }\href {\doibase 10.1038/nmat2166} {\bibfield  {journal} {\bibinfo  {journal}
  {Nature Materials}\ }\textbf {\bibinfo {volume} {7}},\ \bibinfo {pages} {406}
  (\bibinfo {year} {2008})}\BibitemShut {NoStop}%
\bibitem [{\citenamefont {Wang}\ and\ \citenamefont
  {Beasley}(2013)}]{Wang2013}%
  \BibitemOpen
  \bibfield  {author} {\bibinfo {author} {\bibfnamefont {W.}~\bibnamefont
  {Wang}}\ and\ \bibinfo {author} {\bibfnamefont {M.~R.}\ \bibnamefont
  {Beasley}},\ }\href {\doibase 10.1063/1.4796175} {\bibfield  {journal}
  {\bibinfo  {journal} {Applied Physics Letters}\ }\textbf {\bibinfo {volume}
  {102}},\ \bibinfo {pages} {131605} (\bibinfo {year} {2013})}\BibitemShut
  {NoStop}%
\bibitem [{\citenamefont {Buron}\ \emph {et~al.}(2012)\citenamefont {Buron},
  \citenamefont {Petersen}, \citenamefont {B{\o}ggild}, \citenamefont {Cooke},
  \citenamefont {Hilke}, \citenamefont {Sun}, \citenamefont {Whiteway},
  \citenamefont {Nielsen}, \citenamefont {Hansen}, \citenamefont {Yurgens},\
  and\ \citenamefont {Jepsen}}]{Buron2012}%
  \BibitemOpen
  \bibfield  {author} {\bibinfo {author} {\bibfnamefont {J.~D.}\ \bibnamefont
  {Buron}}, \bibinfo {author} {\bibfnamefont {D.~H.}\ \bibnamefont {Petersen}},
  \bibinfo {author} {\bibfnamefont {P.}~\bibnamefont {B{\o}ggild}}, \bibinfo
  {author} {\bibfnamefont {D.~G.}\ \bibnamefont {Cooke}}, \bibinfo {author}
  {\bibfnamefont {M.}~\bibnamefont {Hilke}}, \bibinfo {author} {\bibfnamefont
  {J.}~\bibnamefont {Sun}}, \bibinfo {author} {\bibfnamefont {E.}~\bibnamefont
  {Whiteway}}, \bibinfo {author} {\bibfnamefont {P.~F.}\ \bibnamefont
  {Nielsen}}, \bibinfo {author} {\bibfnamefont {O.}~\bibnamefont {Hansen}},
  \bibinfo {author} {\bibfnamefont {A.}~\bibnamefont {Yurgens}}, \ and\
  \bibinfo {author} {\bibfnamefont {P.~U.}\ \bibnamefont {Jepsen}},\ }\href
  {\doibase 10.1021/nl301551a} {\bibfield  {journal} {\bibinfo  {journal} {Nano
  Letters}\ }\textbf {\bibinfo {volume} {12}},\ \bibinfo {pages} {5074}
  (\bibinfo {year} {2012})}\BibitemShut {NoStop}%
\bibitem [{\citenamefont {Borunda}\ \emph {et~al.}(2013)\citenamefont
  {Borunda}, \citenamefont {Hennig},\ and\ \citenamefont
  {Heller}}]{Borunda2013}%
  \BibitemOpen
  \bibfield  {author} {\bibinfo {author} {\bibfnamefont {M.~F.}\ \bibnamefont
  {Borunda}}, \bibinfo {author} {\bibfnamefont {H.}~\bibnamefont {Hennig}}, \
  and\ \bibinfo {author} {\bibfnamefont {E.~J.}\ \bibnamefont {Heller}},\
  }\href {\doibase 10.1103/PhysRevB.88.125415} {\bibfield  {journal} {\bibinfo
  {journal} {Physical Review B}\ }\textbf {\bibinfo {volume} {88}},\ \bibinfo
  {pages} {125415} (\bibinfo {year} {2013})}\BibitemShut {NoStop}%
\bibitem [{\citenamefont {Mayorov}\ \emph {et~al.}(2011)\citenamefont
  {Mayorov}, \citenamefont {Gorbachev}, \citenamefont {Morozov}, \citenamefont
  {Britnell}, \citenamefont {Jalil}, \citenamefont {Ponomarenko}, \citenamefont
  {Blake}, \citenamefont {Novoselov}, \citenamefont {Watanabe}, \citenamefont
  {Taniguchi},\ and\ \citenamefont {Geim}}]{Mayorov2011}%
  \BibitemOpen
  \bibfield  {author} {\bibinfo {author} {\bibfnamefont {A.~S.}\ \bibnamefont
  {Mayorov}}, \bibinfo {author} {\bibfnamefont {R.~V.}\ \bibnamefont
  {Gorbachev}}, \bibinfo {author} {\bibfnamefont {S.~V.}\ \bibnamefont
  {Morozov}}, \bibinfo {author} {\bibfnamefont {L.}~\bibnamefont {Britnell}},
  \bibinfo {author} {\bibfnamefont {R.}~\bibnamefont {Jalil}}, \bibinfo
  {author} {\bibfnamefont {L.~A.}\ \bibnamefont {Ponomarenko}}, \bibinfo
  {author} {\bibfnamefont {P.}~\bibnamefont {Blake}}, \bibinfo {author}
  {\bibfnamefont {K.~S.}\ \bibnamefont {Novoselov}}, \bibinfo {author}
  {\bibfnamefont {K.}~\bibnamefont {Watanabe}}, \bibinfo {author}
  {\bibfnamefont {T.}~\bibnamefont {Taniguchi}}, \ and\ \bibinfo {author}
  {\bibfnamefont {A.~K.}\ \bibnamefont {Geim}},\ }\href {\doibase
  10.1021/nl200758b} {\bibfield  {journal} {\bibinfo  {journal} {Nano Letters}\
  }\textbf {\bibinfo {volume} {11}},\ \bibinfo {pages} {2396} (\bibinfo {year}
  {2011})}\BibitemShut {NoStop}%
\bibitem [{\citenamefont {Berger}\ \emph {et~al.}(2006)\citenamefont {Berger},
  \citenamefont {Song}, \citenamefont {Li}, \citenamefont {Wu}, \citenamefont
  {Brown}, \citenamefont {Naud}, \citenamefont {Mayou}, \citenamefont {Li},
  \citenamefont {Hass}, \citenamefont {Marchenkov}, \citenamefont {Conrad},
  \citenamefont {First},\ and\ \citenamefont {de~Heer}}]{Berger2006}%
  \BibitemOpen
  \bibfield  {author} {\bibinfo {author} {\bibfnamefont {C.}~\bibnamefont
  {Berger}}, \bibinfo {author} {\bibfnamefont {Z.}~\bibnamefont {Song}},
  \bibinfo {author} {\bibfnamefont {X.}~\bibnamefont {Li}}, \bibinfo {author}
  {\bibfnamefont {X.}~\bibnamefont {Wu}}, \bibinfo {author} {\bibfnamefont
  {N.}~\bibnamefont {Brown}}, \bibinfo {author} {\bibfnamefont
  {C.}~\bibnamefont {Naud}}, \bibinfo {author} {\bibfnamefont {D.}~\bibnamefont
  {Mayou}}, \bibinfo {author} {\bibfnamefont {T.}~\bibnamefont {Li}}, \bibinfo
  {author} {\bibfnamefont {J.}~\bibnamefont {Hass}}, \bibinfo {author}
  {\bibfnamefont {A.~N.}\ \bibnamefont {Marchenkov}}, \bibinfo {author}
  {\bibfnamefont {E.~H.}\ \bibnamefont {Conrad}}, \bibinfo {author}
  {\bibfnamefont {P.~N.}\ \bibnamefont {First}}, \ and\ \bibinfo {author}
  {\bibfnamefont {W.~A.}\ \bibnamefont {de~Heer}},\ }\href {\doibase
  10.1126/science.1125925} {\bibfield  {journal} {\bibinfo  {journal}
  {Science}\ }\textbf {\bibinfo {volume} {312}},\ \bibinfo {pages} {1191}
  (\bibinfo {year} {2006})}\BibitemShut {NoStop}%
\bibitem [{\citenamefont {Bolotin}\ \emph {et~al.}(2008)\citenamefont
  {Bolotin}, \citenamefont {Sikes}, \citenamefont {Jiang}, \citenamefont
  {Klima}, \citenamefont {Fudenberg}, \citenamefont {Hone}, \citenamefont
  {Kim},\ and\ \citenamefont {Stormer}}]{Bolotin2008}%
  \BibitemOpen
  \bibfield  {author} {\bibinfo {author} {\bibfnamefont {K.}~\bibnamefont
  {Bolotin}}, \bibinfo {author} {\bibfnamefont {K.}~\bibnamefont {Sikes}},
  \bibinfo {author} {\bibfnamefont {Z.}~\bibnamefont {Jiang}}, \bibinfo
  {author} {\bibfnamefont {M.}~\bibnamefont {Klima}}, \bibinfo {author}
  {\bibfnamefont {G.}~\bibnamefont {Fudenberg}}, \bibinfo {author}
  {\bibfnamefont {J.}~\bibnamefont {Hone}}, \bibinfo {author} {\bibfnamefont
  {P.}~\bibnamefont {Kim}}, \ and\ \bibinfo {author} {\bibfnamefont
  {H.}~\bibnamefont {Stormer}},\ }\href
  {http://www.sciencedirect.com/science/article/pii/S0038109808001178}
  {\bibfield  {journal} {\bibinfo  {journal} {Solid State Communications}\
  }\textbf {\bibinfo {volume} {146}},\ \bibinfo {pages} {351} (\bibinfo {year}
  {2008})}\BibitemShut {NoStop}%
\bibitem [{\citenamefont {Rickhaus}\ \emph {et~al.}(2013)\citenamefont
  {Rickhaus}, \citenamefont {Maurand}, \citenamefont {Liu}, \citenamefont
  {Weiss}, \citenamefont {Richter},\ and\ \citenamefont
  {Sch\"{o}nenberger}}]{Rickhaus2013}%
  \BibitemOpen
  \bibfield  {author} {\bibinfo {author} {\bibfnamefont {P.}~\bibnamefont
  {Rickhaus}}, \bibinfo {author} {\bibfnamefont {R.}~\bibnamefont {Maurand}},
  \bibinfo {author} {\bibfnamefont {M.-H.}\ \bibnamefont {Liu}}, \bibinfo
  {author} {\bibfnamefont {M.}~\bibnamefont {Weiss}}, \bibinfo {author}
  {\bibfnamefont {K.}~\bibnamefont {Richter}}, \ and\ \bibinfo {author}
  {\bibfnamefont {C.}~\bibnamefont {Sch\"{o}nenberger}},\ }\href {\doibase
  10.1038/ncomms3342} {\bibfield  {journal} {\bibinfo  {journal} {Nature
  communications}\ }\textbf {\bibinfo {volume} {4}},\ \bibinfo {pages} {2342}
  (\bibinfo {year} {2013})}\BibitemShut {NoStop}%
\bibitem [{\citenamefont {Baringhaus}\ \emph {et~al.}(2013)\citenamefont
  {Baringhaus}, \citenamefont {Edler},\ and\ \citenamefont
  {Tegenkamp}}]{Baringhaus2013}%
  \BibitemOpen
  \bibfield  {author} {\bibinfo {author} {\bibfnamefont {J.}~\bibnamefont
  {Baringhaus}}, \bibinfo {author} {\bibfnamefont {F.}~\bibnamefont {Edler}}, \
  and\ \bibinfo {author} {\bibfnamefont {C.}~\bibnamefont {Tegenkamp}},\ }\href
  {\doibase 10.1088/0953-8984/25/39/392001} {\bibfield  {journal} {\bibinfo
  {journal} {Journal of Physics: Condensed Matter}\ }\textbf {\bibinfo {volume}
  {25}},\ \bibinfo {pages} {392001} (\bibinfo {year} {2013})}\BibitemShut
  {NoStop}%
\bibitem [{\citenamefont {Eder}\ \emph {et~al.}(2013)\citenamefont {Eder},
  \citenamefont {Kotakoski}, \citenamefont {Holzweber}, \citenamefont
  {Mangler}, \citenamefont {Skakalova},\ and\ \citenamefont
  {Meyer}}]{Eder2013}%
  \BibitemOpen
  \bibfield  {author} {\bibinfo {author} {\bibfnamefont {F.~R.}\ \bibnamefont
  {Eder}}, \bibinfo {author} {\bibfnamefont {J.}~\bibnamefont {Kotakoski}},
  \bibinfo {author} {\bibfnamefont {K.}~\bibnamefont {Holzweber}}, \bibinfo
  {author} {\bibfnamefont {C.}~\bibnamefont {Mangler}}, \bibinfo {author}
  {\bibfnamefont {V.}~\bibnamefont {Skakalova}}, \ and\ \bibinfo {author}
  {\bibfnamefont {J.~C.}\ \bibnamefont {Meyer}},\ }\href {\doibase
  10.1021/nl3042799} {\bibfield  {journal} {\bibinfo  {journal} {Nano Letters}\
  }\textbf {\bibinfo {volume} {13}},\ \bibinfo {pages} {1934} (\bibinfo {year}
  {2013})}\BibitemShut {NoStop}%
\bibitem [{\citenamefont {Datta}(1997)}]{DattaBook}%
  \BibitemOpen
  \bibfield  {author} {\bibinfo {author} {\bibfnamefont {S.}~\bibnamefont
  {Datta}},\ }\href@noop {} {\emph {\bibinfo {title} {Electronic Transport in
  Mesoscopic Systems}}}\ (\bibinfo  {publisher} {Cambridge University Press},\
  \bibinfo {year} {1997})\BibitemShut {NoStop}%
\bibitem [{\citenamefont {Haug}\ and\ \citenamefont {Jauho}(2008)}]{AnttiBook}%
  \BibitemOpen
  \bibfield  {author} {\bibinfo {author} {\bibfnamefont {H.}~\bibnamefont
  {Haug}}\ and\ \bibinfo {author} {\bibfnamefont {A.-P.}\ \bibnamefont
  {Jauho}},\ }\href@noop {} {\emph {\bibinfo {title} {Quantum kinetics in
  transport and optics of semiconductors}}}\ (\bibinfo  {publisher}
  {Springer},\ \bibinfo {year} {2008})\BibitemShut {NoStop}%
\bibitem [{\citenamefont {Power}\ and\ \citenamefont
  {Ferreira}(2011)}]{Power2011}%
  \BibitemOpen
  \bibfield  {author} {\bibinfo {author} {\bibfnamefont {S.~R.}\ \bibnamefont
  {Power}}\ and\ \bibinfo {author} {\bibfnamefont {M.~S.}\ \bibnamefont
  {Ferreira}},\ }\href {\doibase 10.1103/PhysRevB.83.155432} {\bibfield
  {journal} {\bibinfo  {journal} {Physical Review B}\ }\textbf {\bibinfo
  {volume} {83}},\ \bibinfo {pages} {155432} (\bibinfo {year}
  {2011})}\BibitemShut {NoStop}%
\bibitem [{\citenamefont {Reich}\ \emph {et~al.}(2002)\citenamefont {Reich},
  \citenamefont {Maultzsch}, \citenamefont {Thomsen},\ and\ \citenamefont
  {Ordej\'{o}n}}]{Reich2002}%
  \BibitemOpen
  \bibfield  {author} {\bibinfo {author} {\bibfnamefont {S.}~\bibnamefont
  {Reich}}, \bibinfo {author} {\bibfnamefont {J.}~\bibnamefont {Maultzsch}},
  \bibinfo {author} {\bibfnamefont {C.}~\bibnamefont {Thomsen}}, \ and\
  \bibinfo {author} {\bibfnamefont {P.}~\bibnamefont {Ordej\'{o}n}},\ }\href
  {\doibase 10.1103/PhysRevB.66.035412} {\bibfield  {journal} {\bibinfo
  {journal} {Physical Review B}\ }\textbf {\bibinfo {volume} {66}},\ \bibinfo
  {pages} {035412} (\bibinfo {year} {2002})}\BibitemShut {NoStop}%
\bibitem [{\citenamefont {Meunier}\ and\ \citenamefont
  {Lambin}(1998)}]{Meunier1998}%
  \BibitemOpen
  \bibfield  {author} {\bibinfo {author} {\bibfnamefont {V.}~\bibnamefont
  {Meunier}}\ and\ \bibinfo {author} {\bibfnamefont {P.}~\bibnamefont
  {Lambin}},\ }\href {\doibase 10.1103/PhysRevLett.81.5588} {\bibfield
  {journal} {\bibinfo  {journal} {Physical Review Letters}\ }\textbf {\bibinfo
  {volume} {81}},\ \bibinfo {pages} {5588} (\bibinfo {year}
  {1998})}\BibitemShut {NoStop}%
\bibitem [{\citenamefont {Fukuda}\ \emph {et~al.}(2007)\citenamefont {Fukuda},
  \citenamefont {Oymak},\ and\ \citenamefont {Hong}}]{Fukuda2007}%
  \BibitemOpen
  \bibfield  {author} {\bibinfo {author} {\bibfnamefont {T.}~\bibnamefont
  {Fukuda}}, \bibinfo {author} {\bibfnamefont {H.}~\bibnamefont {Oymak}}, \
  and\ \bibinfo {author} {\bibfnamefont {J.}~\bibnamefont {Hong}},\ }\href
  {\doibase 10.1103/PhysRevB.75.195428} {\bibfield  {journal} {\bibinfo
  {journal} {Physical Review B}\ }\textbf {\bibinfo {volume} {75}},\ \bibinfo
  {pages} {195428} (\bibinfo {year} {2007})}\BibitemShut {NoStop}%
\bibitem [{\citenamefont {Nakanishi}\ and\ \citenamefont
  {Ando}(2010)}]{Nakanishi2010}%
  \BibitemOpen
  \bibfield  {author} {\bibinfo {author} {\bibfnamefont {T.}~\bibnamefont
  {Nakanishi}}\ and\ \bibinfo {author} {\bibfnamefont {T.}~\bibnamefont
  {Ando}},\ }\href {\doibase http://dx.doi.org/10.1016/j.physe.2009.10.041}
  {\bibfield  {journal} {\bibinfo  {journal} {Physica E: Low-dimensional
  Systems and Nanostructures}\ }\textbf {\bibinfo {volume} {42}},\ \bibinfo
  {pages} {726} (\bibinfo {year} {2010})}\BibitemShut {NoStop}%
\bibitem [{Note1()}]{Note1}%
  \BibitemOpen
  \bibinfo {note} {The coupling is calculated using\cite
  {Meunier1998,Amara2007}, $t_i = t_0 w_i \protect \mathrm {e}^{-d_i/\lambda
  }\protect \qopname \relax o{cos}{\setbox \z@ \hbox {\frozen@everymath
  \@emptytoks \mathsurround \z@ $\nulldelimiterspace \z@ \left (\vcenter to\@ne
  \big@size {}\right .$}\box \z@ }\theta _i{\setbox \z@ \hbox
  {\frozen@everymath \@emptytoks \mathsurround \z@ $\nulldelimiterspace \z@
  \left )\vcenter to\@ne \big@size {}\right .$}\box \z@ }$ where $w_i =
  \protect \mathrm {e}^{-ad_i^2}/\DOTSB \sum@ \slimits@ _{j} \protect \mathrm
  {e}^{-ad_j^2}$, $\theta _i$ is the angle between the tip apex and site $i$,
  $\lambda =0.85\r A$, $a=0.6 \r A^{-2}$. $t_0$ is a scaling factor which we
  set to $t_0=10t$.}\BibitemShut {Stop}%
\bibitem [{\citenamefont {Ribeiro}\ \emph {et~al.}(2009)\citenamefont
  {Ribeiro}, \citenamefont {Pereira}, \citenamefont {Peres}, \citenamefont
  {Briddon},\ and\ \citenamefont {{Castro Neto}}}]{Ribeiro2009}%
  \BibitemOpen
  \bibfield  {author} {\bibinfo {author} {\bibfnamefont {R.~M.}\ \bibnamefont
  {Ribeiro}}, \bibinfo {author} {\bibfnamefont {V.~M.}\ \bibnamefont
  {Pereira}}, \bibinfo {author} {\bibfnamefont {N.~M.~R.}\ \bibnamefont
  {Peres}}, \bibinfo {author} {\bibfnamefont {P.~R.}\ \bibnamefont {Briddon}},
  \ and\ \bibinfo {author} {\bibfnamefont {A.~H.}\ \bibnamefont {{Castro
  Neto}}},\ }\href {\doibase 10.1088/1367-2630/11/11/115002} {\bibfield
  {journal} {\bibinfo  {journal} {New Journal of Physics}\ }\textbf {\bibinfo
  {volume} {11}},\ \bibinfo {pages} {115002} (\bibinfo {year}
  {2009})}\BibitemShut {NoStop}%
\bibitem [{\citenamefont {Dubois}\ \emph {et~al.}(2010)\citenamefont {Dubois},
  \citenamefont {Lopez-Bezanilla}, \citenamefont {Cresti}, \citenamefont
  {Triozon}, \citenamefont {Biel}, \citenamefont {Charlier},\ and\
  \citenamefont {Roche}}]{Dubois2010}%
  \BibitemOpen
  \bibfield  {author} {\bibinfo {author} {\bibfnamefont {S.~M.-M.}\
  \bibnamefont {Dubois}}, \bibinfo {author} {\bibfnamefont {A.}~\bibnamefont
  {Lopez-Bezanilla}}, \bibinfo {author} {\bibfnamefont {A.}~\bibnamefont
  {Cresti}}, \bibinfo {author} {\bibfnamefont {F.}~\bibnamefont {Triozon}},
  \bibinfo {author} {\bibfnamefont {B.}~\bibnamefont {Biel}}, \bibinfo {author}
  {\bibfnamefont {J.-C.}\ \bibnamefont {Charlier}}, \ and\ \bibinfo {author}
  {\bibfnamefont {S.}~\bibnamefont {Roche}},\ }\href {\doibase
  10.1021/nn100028q} {\bibfield  {journal} {\bibinfo  {journal} {ACS nano}\
  }\textbf {\bibinfo {volume} {4}},\ \bibinfo {pages} {1971} (\bibinfo {year}
  {2010})}\BibitemShut {NoStop}%
\bibitem [{\citenamefont {Wakabayashi}\ \emph {et~al.}(2007)\citenamefont
  {Wakabayashi}, \citenamefont {Takane},\ and\ \citenamefont
  {Sigrist}}]{Wakabayashi2007}%
  \BibitemOpen
  \bibfield  {author} {\bibinfo {author} {\bibfnamefont {K.}~\bibnamefont
  {Wakabayashi}}, \bibinfo {author} {\bibfnamefont {Y.}~\bibnamefont {Takane}},
  \ and\ \bibinfo {author} {\bibfnamefont {M.}~\bibnamefont {Sigrist}},\ }\href
  {\doibase 10.1103/PhysRevLett.99.036601} {\bibfield  {journal} {\bibinfo
  {journal} {Physical Review Letters}\ }\textbf {\bibinfo {volume} {99}},\
  \bibinfo {pages} {036601} (\bibinfo {year} {2007})}\BibitemShut {NoStop}%
\bibitem [{Note2()}]{Note2}%
  \BibitemOpen
  \bibinfo {note} {The GF including an armchair edge, is calculated from the
  pristine GF with the method of images, as described in Ref. \cite
  {Duffy2013}. For the zigzag edge a direct inversion scheme is
  used.}\BibitemShut {Stop}%
\bibitem [{\citenamefont {Casiraghi}\ \emph {et~al.}(2009)\citenamefont
  {Casiraghi}, \citenamefont {Hartschuh}, \citenamefont {Qian}, \citenamefont
  {Piscanec}, \citenamefont {Georgi}, \citenamefont {Fasoli}, \citenamefont
  {Novoselov}, \citenamefont {Basko},\ and\ \citenamefont
  {Ferrari}}]{Casiraghi2009}%
  \BibitemOpen
  \bibfield  {author} {\bibinfo {author} {\bibfnamefont {C.}~\bibnamefont
  {Casiraghi}}, \bibinfo {author} {\bibfnamefont {A.}~\bibnamefont
  {Hartschuh}}, \bibinfo {author} {\bibfnamefont {H.}~\bibnamefont {Qian}},
  \bibinfo {author} {\bibfnamefont {S.}~\bibnamefont {Piscanec}}, \bibinfo
  {author} {\bibfnamefont {C.}~\bibnamefont {Georgi}}, \bibinfo {author}
  {\bibfnamefont {A.}~\bibnamefont {Fasoli}}, \bibinfo {author} {\bibfnamefont
  {K.~S.}\ \bibnamefont {Novoselov}}, \bibinfo {author} {\bibfnamefont {D.~M.}\
  \bibnamefont {Basko}}, \ and\ \bibinfo {author} {\bibfnamefont {A.~C.}\
  \bibnamefont {Ferrari}},\ }\href {\doibase 10.1021/nl8032697} {\bibfield
  {journal} {\bibinfo  {journal} {Nano Letters}\ }\textbf {\bibinfo {volume}
  {9}},\ \bibinfo {pages} {1433} (\bibinfo {year} {2009})}\BibitemShut
  {NoStop}%
\bibitem [{Note3()}]{Note3}%
  \BibitemOpen
  \bibinfo {note} {Random smooth height fluctuations with amplitudes in
  accordance to \cite {Fasolino2007}. The electronic effects can be accounted
  for by varying the hopping parameters according to \cite {Pereira2009} as
  $t=t_0\protect \mathrm {e}^{-3.37 (a/a_0-1)}$, where $a$ is the new bond
  length and $a_0=1.42 \r A$.}\BibitemShut {Stop}%
\bibitem [{\citenamefont {Juan}\ \emph {et~al.}(2011)\citenamefont {Juan},
  \citenamefont {Cortijo}, \citenamefont {Vozmediano},\ and\ \citenamefont
  {Cano}}]{Juan2011}%
  \BibitemOpen
  \bibfield  {author} {\bibinfo {author} {\bibfnamefont {F.~D.}\ \bibnamefont
  {Juan}}, \bibinfo {author} {\bibfnamefont {A.}~\bibnamefont {Cortijo}},
  \bibinfo {author} {\bibfnamefont {M.~A.~H.}\ \bibnamefont {Vozmediano}}, \
  and\ \bibinfo {author} {\bibfnamefont {A.}~\bibnamefont {Cano}},\ }\href
  {\doibase 10.1038/nphys2034} {\bibfield  {journal} {\bibinfo  {journal}
  {Nature Physics}\ }\textbf {\bibinfo {volume} {7}},\ \bibinfo {pages} {810}
  (\bibinfo {year} {2011})}\BibitemShut {NoStop}%
\bibitem [{\citenamefont {Duffy}\ \emph {et~al.}(2014)\citenamefont {Duffy},
  \citenamefont {Gorman}, \citenamefont {Power},\ and\ \citenamefont
  {Ferreira}}]{Duffy2013}%
  \BibitemOpen
  \bibfield  {author} {\bibinfo {author} {\bibfnamefont {J.~M.}\ \bibnamefont
  {Duffy}}, \bibinfo {author} {\bibfnamefont {P.~D.}\ \bibnamefont {Gorman}},
  \bibinfo {author} {\bibfnamefont {S.~R.}\ \bibnamefont {Power}}, \ and\
  \bibinfo {author} {\bibfnamefont {M.~S.}\ \bibnamefont {Ferreira}},\ }\href
  {http://stacks.iop.org/0953-8984/26/i=5/a=055007} {\bibfield  {journal}
  {\bibinfo  {journal} {Journal of Physics: Condensed Matter}\ }\textbf
  {\bibinfo {volume} {26}},\ \bibinfo {pages} {055007} (\bibinfo {year}
  {2014})}\BibitemShut {NoStop}%
\bibitem [{\citenamefont {Fasolino}\ \emph {et~al.}(2007)\citenamefont
  {Fasolino}, \citenamefont {Los},\ and\ \citenamefont
  {Katsnelson}}]{Fasolino2007}%
  \BibitemOpen
  \bibfield  {author} {\bibinfo {author} {\bibfnamefont {A.}~\bibnamefont
  {Fasolino}}, \bibinfo {author} {\bibfnamefont {J.~H.}\ \bibnamefont {Los}}, \
  and\ \bibinfo {author} {\bibfnamefont {M.~I.}\ \bibnamefont {Katsnelson}},\
  }\href {\doibase 10.1038/nmat2011} {\bibfield  {journal} {\bibinfo  {journal}
  {Nature Materials}\ }\textbf {\bibinfo {volume} {6}},\ \bibinfo {pages} {858}
  (\bibinfo {year} {2007})}\BibitemShut {NoStop}%
\bibitem [{\citenamefont {Pereira}\ \emph {et~al.}(2009)\citenamefont
  {Pereira}, \citenamefont {{Castro Neto}},\ and\ \citenamefont
  {Peres}}]{Pereira2009}%
  \BibitemOpen
  \bibfield  {author} {\bibinfo {author} {\bibfnamefont {V.}~\bibnamefont
  {Pereira}}, \bibinfo {author} {\bibfnamefont {A.~H.}\ \bibnamefont {{Castro
  Neto}}}, \ and\ \bibinfo {author} {\bibfnamefont {N.~M.~R.}\ \bibnamefont
  {Peres}},\ }\href {\doibase 10.1103/PhysRevB.80.045401} {\bibfield  {journal}
  {\bibinfo  {journal} {Physical Review B}\ }\textbf {\bibinfo {volume} {80}},\
  \bibinfo {pages} {045401} (\bibinfo {year} {2009})}\BibitemShut {NoStop}%
\end{thebibliography}
%

\end{document}